\documentclass[twocolumn]{aastex62}
\usepackage{epsfig}
\usepackage{graphicx}
\usepackage{url}

\begin{document}

\title{The Nearby, Young, $\chi$$^1$ Fornacis Cluster: Membership, Age, and an Extraordinary Ensemble of Dusty Debris Disks}

% AUTHOR AND ADDRESS FORMAT FOR IOPART CLASS FILE:
%\author{B. Zuckerman$^1$, Beth Klein$^1$, Joel Kastner$^2$}
%\address{$^1$Department of Physics and Astronomy, University of California, Los Angeles, CA 90095, USA}
%\address{$^2$Chester F. Carlson Center for Imaging Science, School of Physics and Astronomy, \& Laboratory for Multiwavelength Astrophysics, Rochester Institute of Technology, Rochester, NY 14623, USA}
%\eads{\mailto{ben@astro.ucla.edu}}

% AUTHOR AND ADDRESS FORMAT FOR AASTEX62 CLASS FILE:
\correspondingauthor{B. Zuckerman}
\email{ben@astro.ucla.edu}

\author{B. Zuckerman}
\affiliation{Department of Physics and Astronomy, University of California, Los Angeles, CA 90095-1562, USA}

\author{Beth Klein}
\affiliation{Department of Physics and Astronomy, University of California, Los Angeles, CA 90095-1562, USA}

\author{Joel Kastner}
\affiliation{Chester F. Carlson Center for Imaging Science, School of Physics and Astronomy, \& Laboratory for Multiwavelength Astrophysics, Rochester Institute of Technology, Rochester, NY 14623, USA}

\begin{abstract}

Only four star clusters are known within $\sim$100 pc of Earth.  Of these, 
the $\chi$$^1$ For cluster has barely been studied.  We use the {\it Gaia} DR2 
catalog and other published data to establish the cluster membership, 
structure, and age. The age of and distance to the cluster are $\sim$40 Myr and 104 pc, respectively.
A remarkable, unprecedented, aspect of the 
cluster is the large percentage of M-type stars with warm excess 
infrared emission due to orbiting dust grains $-$ these stars lie in 
an annulus that straddles the tidal radius of the cluster.  The 
$\chi$$^1$ For cluster appears to be closely related to two extensive, 
previously known, groups of comoving, coeval stars (the Tucana-Horologium 
and Columba Associations) that are spread over much of the southern sky.  
While Tuc-Hor and $\chi$$^1$ For are comoving and coeval, the difference in the frequency of
their warm dusty debris disks at M-type stars could hardly be more dramatic.

\end{abstract}
%\pacs{97.10.Tk}
%\maketitle

\section{Introduction} 
Star clusters have been recognized since antiquity $-$ the Pleiades and the 
Hyades can be seen with the naked eye $-$ and have been carefully studied by 
astronomers since the invention of the telescope.  Both open and globular 
clusters have played an essential role in our understanding of stellar 
evolution.  Measurements by the {\it Gaia} satellite of the European Space Agency 
have enabled major advances in recognition and characterization of Galactic 
star clusters (e.g., Cantat-Gaudin et al 2018; Lodieu et al 2019).  To date, 
{\it {\it Gaia}} findings have been reported mostly as general overviews of cluster 
existence and of some cluster properties rather than as detailed 
investigations of specific clusters.  In the present paper, we use {\it Gaia} 
along with additional information to call attention to the significance of a 
nearby, but previously little studied open cluster initially named Alessi 13 
(Dias et al 2002).

G. Cutispoto and colleagues, in a series of papers in the decade of the 1990s, focused on stellar activity, including X-rays and lithium abundances, to identify some young stars near Earth.  But kinematic moving groups were not clearly recognized and the papers were poorly cited.  de la Reza et al (1989) and Gregorio-Hetem et al (1992) identified a few late-type stars
near the infrared bright star TW Hya that also exhibited spectral 
signatures of youth, i.e., strong Li absorption, H-alpha emission, 
and/or IR excess.   The arguably first 
paper to provide some order to this new field involved this group of $\sim$10 Myr old 
stars $-$ the TW Hya Association $-$ located about 60 pc from Earth (Kastner 
et al 1997).  Since then perhaps as many as 10 other associations of stars 
with ages 200 Myr or less and within $\sim$100 pc of Earth have been identified 
(e.g., Zuckerman \& Song 2004; Torres et al 2008; Mamajek 2016; Gagn{\'e} et al 
2018).  Akin to star clusters, members of a given nearby, young stellar 
association or moving group have similar ages and space motions through the Milky 
Way (they are coeval and comoving).  But unlike star clusters, when 
considered in their entirety, stars in the various nearby young associations 
are distributed over much of the sky.  Studies of these associations are 
enabling astronomers to better understand the evolution of stars as they age 
from 10 to 200 Myr.  In addition, and perhaps most important, these youthful 
stars have provided numerous targets for direct infrared imaging discovery 
of extrasolar planets.

\begin{figure*}
\plotone{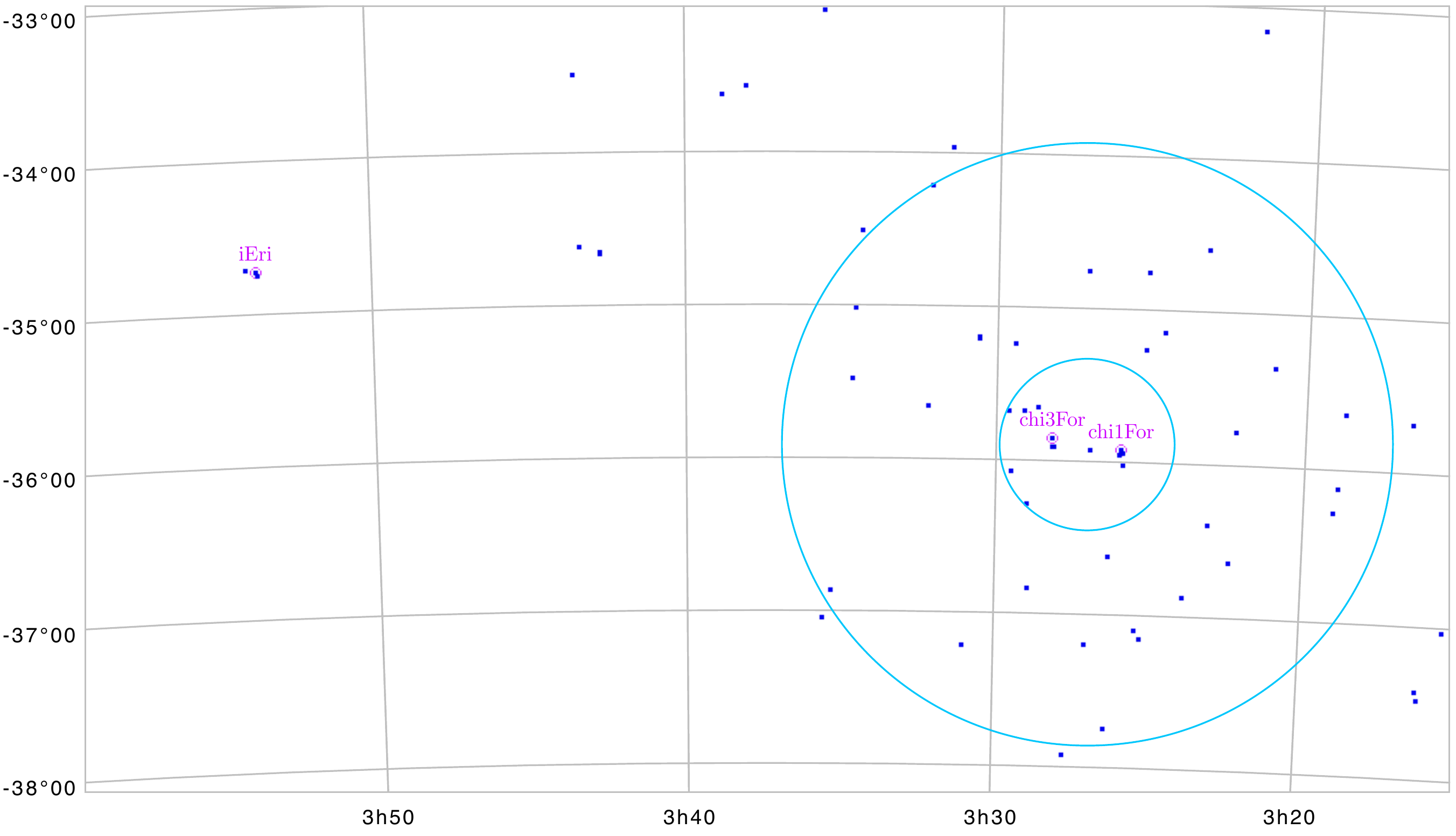}
\caption{Sky plot of stars of the $\chi$$^1$ For cluster. As discussed in the text, we assume the cluster center to lie midway between the stars $\chi$$^1$ For and $\chi$$^3$ For. The small and large circles indicate the cluster's core and tidal radii, respectively (see Section 4). 
%Some of the plotted stars that appear to be within the tidal radius, are actually  
The luminous star i Eri (mentioned in the text) is indicated.}
\label{fig:skyplot}
\end{figure*}

Following Dias et al (2002), Yen et al (2018) used TGAS ({\it Gaia} DR1) to analyze 24 nearby open clusters including Alessi 13.   Yen et al identified nine stars in TGAS as cluster members.   Using {\it Gaia} DR2, Cantat-Gaudin et al. (2018) analyzed 1229 open clusters and reported 48 candidate members of Alessi 13 contained in a volume of radius 15 pc.   In the present paper we use {\it Gaia} DR2 to greatly expand on the known population and properties of Alessi 13.  In particular, by choosing a less restrictive range of cluster kinematic properties we have identified 108 candidate members within a similar volume of radius 15 pc. 
Our list includes 47 of the 48 stars listed in Cantat-Gaudin et al. 

We demonstrate that the Alessi 13 cluster and two of the 
most prominent of the nearby youthful stellar associations $-$ 
Tucana-Horologium and Columba $-$ are coeval and comoving with age $\sim$40 
Myr. Taken together, they constitute a major constituent of the solar 
neighborhood.  For reasons given in Section 4, henceforth we call Alessi 13 the 
$\chi$$^1$ For cluster. $\chi$$^1$ For is the second youngest of only 
four star clusters known within $\sim$100 pc of Earth (the other three are 
mentioned in Section 4).

A substantial fraction of stars of solar mass and above are known to be 
orbited by dusty debris disks, most of which are more akin to the Sun's 
Kuiper Belt than to its asteroid belt.  However, very few low-mass stars are 
known that harbor debris disks $-$ especially when considered as a tiny 
fraction of the population of the numerous known M-type stars.  
As we describe in Section 4, a (currently) unique aspect of the $\chi$$^1$ For 
cluster is its abundance of M-type stars that are orbited by dust particles.  
Perhaps as noteworthy is the large percentage of these debris disks 
that contain warm dust with temperatures characteristic of the asteroid belt 
or of the habitable zone, in contrast to the cold dust of the Kuiper Belt.  
This warm dust may be associated with copious rocky planet formation at a 
characteristic age of $\sim$40 Myr (Melis et al 2010).

\section{Sample selection} \label{sec:selection}

Our path to the $\chi$$^1$ For cluster began with our study of stars in the Columba Association (Lee \& Song 2019).  Our study includes investigation of binary members of this association and we were surprised to discover that one of the Lee-Song stars is apparently surrounded by numerous ``companions''.  One thing led to another, including our discovery of the Dias et al (2002), Yen et al (2018), and Cantat-Gaudin et al. (2018) papers, as the existence of the cluster became apparent to us.   We used these papers as initial guidelines to develop our own set of search constraints, as described below.

\begin{figure}
\plotone{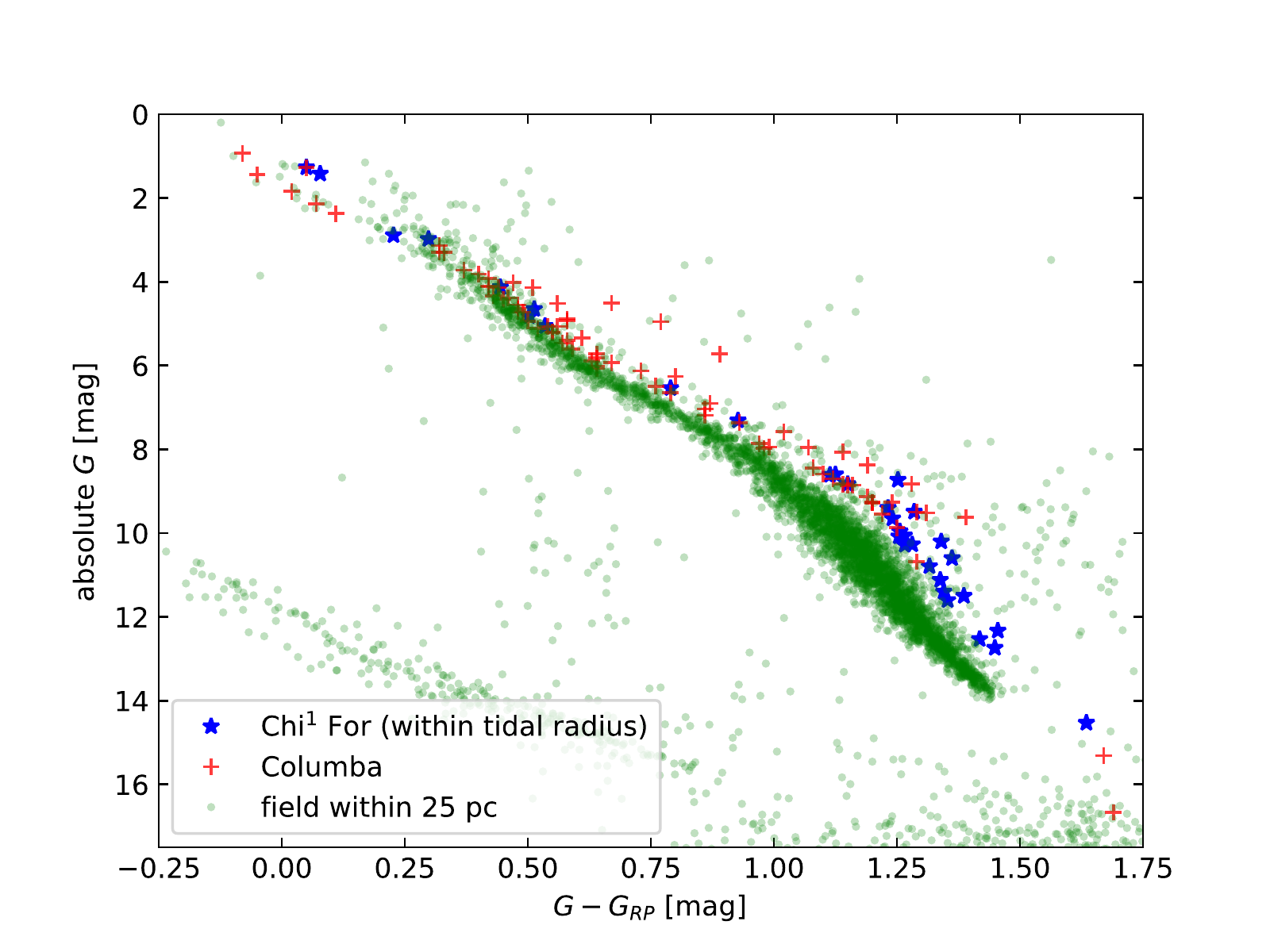}
\caption{Plot of $\chi$$^1$ For stars listed in Tables 1 \& 2
and of quality 1 (highest confidence) Columba member stars given in Lee \& Song (2019). The ordinate is absolute {\it Gaia} G magnitude and the abscissa is {\it Gaia} $G-G_{RP}$ color; these are proxies for stellar luminosity and temperature, respectively (with luminosity increasing as absolute $G$ decreases, and temperature increasing to the left). For comparison, field stars within 25 pc of Earth are also plotted.}
\label{fig:cmd}
\end{figure}

Based on the clustering of massive stars, and the fact that previous papers had identified 
the cluster center in this vicinity, we took the average of the {\it Gaia} positions 
of $\chi$$^1$ and $\chi$$^3$ For as defining the center of the cluster 
at R.A. = 51.766 deg, decl. = -35.887 deg.  We note that this is in good
agreement with the Cantat-Gaudin et al (2018) average of 48 stars at 51.762, -35.821  
(found by starting their search at the Dias et al 2002 listed 
center of 50.425, -35.700, about 1.3 deg West of the new cluster center). 
In the current paper, from our broad search (described below) we find the average position of the cluster core stars to be 51.758, -35.922, close to the average position of the two A1 stars $\chi$$^1$ and $\chi$$^3$ For (Figure 1).

We use the {\it Gaia} DR2 catalog ({\it Gaia} Collaboration 2018; Lindegren et al 2018) 
to identify members of the $\chi$$^1$ For cluster (Tables 1 
\& 2).  Stars in the tables are deemed to be cluster members based on their 
proper motions, parallax, and location on a color-magnitude diagram (CMD, Figure \ref{fig:cmd}).   
Based on the radial extent of cluster members identified by Cantat-Gaudin et al, and 
our concern that extending our search too far from the cluster center would
cause inclusion of or confusion with Columba and Tuc-Hor stars, we searched for stars 
that have R.A. and decl. positions within a 15 pc 
(8.24 degree angular) radius of our adopted cluster center at the mean position 
of the stars $\chi$$^1$ and $\chi$$^3$ For. We selected stars that 
satisfied constraints in parallax and proper motion consistent with being a 
comoving group in a volume of space of $\sim$15 pc radial extent. That is, for
a 15 pc search radius in the R.A.-decl. plane of the sky, we included a range in 
parallax that corresponds
to a $\pm$15 pc line-of-sight distance range, centered on the mean distance 
of $\chi$$^1$ and $\chi$$^3$ For from the Sun (104 pc), and then we
widened that range to allow for parallax uncertainties up to 0.4 mas.

\begin{figure}
\plotone{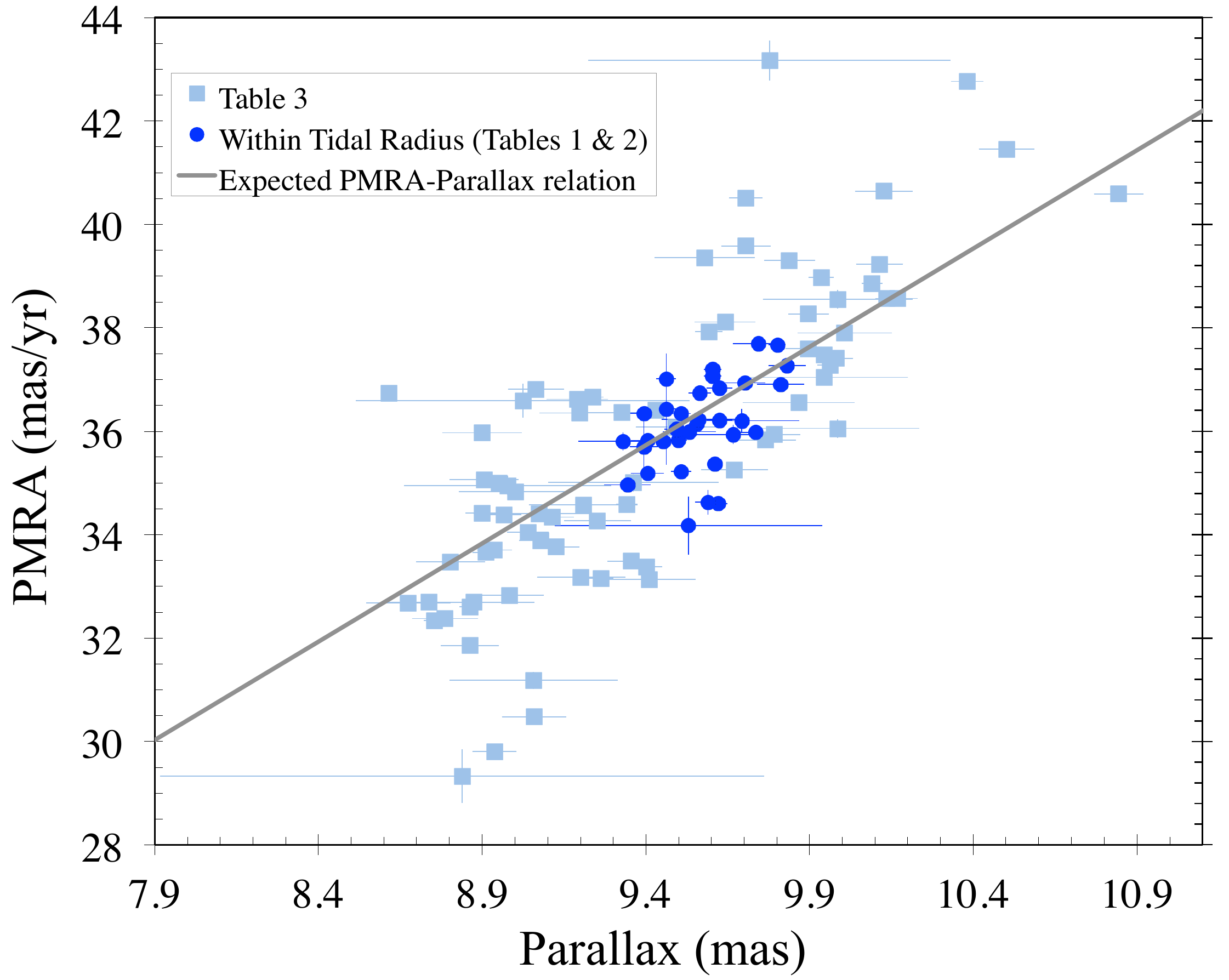}
\caption{{\it Gaia} values for proper motion in R.A. (PMRA) and parallax for $\chi$$^1$ For cluster members and candidates.  As expected for a cluster, due to projection effects based on distance from the Sun, stars with similar space velocities will display apparent proper motions that vary in proportion to their parallax (i.e. more distant stars will appear to have smaller motions). The gray/slanted line indicates this proportionality relation with respect to the average of the proper motions measured for the cluster center members, $\chi$$^1$ For and $\chi$$^3$ For. The error bars are directly from {\it Gaia}; some are smaller than the size of the data points.}
\label{fig:pmra-vs-parallax}
\end{figure}

Next we estimated the corresponding range in expected proper motion values such that it includes the
anticipated differences in proper motion due to projection effects at different distances
from the Sun (Figure \ref{fig:pmra-vs-parallax}).  
%surrounding the average of the proper motions measured for the defining core members, $\chi$$^1$ For and $\chi$$^3$ For = 36.47 $\pm$ 0.08 mas/yr
Note that in the case of the $\chi$$^1$ For cluster, this effect is mainly 
present in only the R.A. proper motion (pmra) direction
which dominates the overall proper motion vector, as the decl. proper motion (pmdec) values are close to zero.  Not surprisingly, stars within the tidal radius (dark-blue/round points) define a compact central region in pmra-parallax space. Some of the light-blue/square points (from Table 3) that plot far from the gray/slanted line may be members of the Tuc-Hor or Columba associations and not of the cluster. 

Thus, the search values for pmra were calculated as centered on the average of the pmra values of the core stars $\chi$$^1$ and $\chi$$^3$ For (36.5 mas/yr) and scaled to the limits of the parallax constraints, resulting in a $+$7.6/$-$6.0 mas/yr range.  Again, we widened that range to allow for pmra uncertainties up to 1 mas/yr.  Since pmdec is close to zero, parallax scaling results in too small of a search space, so for pmdec we adopt a similar range of parallax scaling as found for pmra $-$  i.e., $\pm$7 mas/yr $-$ relative to the average of the pmdec values of the core stars $\chi$$^1$ and $\chi$$^3$ For (-4.2 mas/yr).

With the above considerations we adopted the following {\it Gaia} search constraints: 

\begin{center}
8.0 mas $<$ parallax $<$ 11.6 mas \\   
29 mas/yr $<$ pmra  $<$  45 mas/yr \\
-11 mas/yr $<$ pmdec $<$ 3 mas/yr \\
\end{center}

In a few cases we had to modify values of some {\it Gaia} DR2 entries.  For 
example, {\it Gaia} often has trouble accurately measuring parallax and/or proper 
motions of stars in (moderately) close binaries.  If the binary members are 
of substantially unequal brightness, then the errors listed in the catalog 
for these quantities for the companion are usually substantially larger than those
for the primary. Then, typically, we adopt the parameters of the primary for 
both stars.  If the binary members are of more nearly equal brightness, 
then, for both stars, we adopt an error-weighted average of the relevant 
parameters.

Also, there are occasional errors in {\it Gaia} DR2 for reasons that are not easy 
to discern.  One representative 
example is the proper motion of the star CD-36 1289 which is an obvious 
member of the core of the $\chi$$^1$ For cluster (see Table 1).  The {\it Gaia} 
R.A. proper motion of CD-36 1289 is given as 25 mas/yr.  If this 
were to be correct, then CD-36 1289 could not be a member of the cluster 
because it would have the wrong Galactic space motions UVW (see Section 4).  
However, if one instead uses the well-measured proper motion of 34.6 mas/yr 
from the UCAC4 catalog, then the UVW of CD-36 1289 agrees well with that of 
other cluster core members.  The UCAC4 proper motion agrees well with the 
independent measurements from SPM4, UCAC5, and PPMXL, and thus is obviously 
correct.  (All these catalog values can be found in Vizier.\footnote{\url{http://vizier.u-strasbg.fr/viz-bin/VizieR}})

Our {\it Gaia} catalog search using the above parameters returns 142 sources, plus we add CD-36 1289, 
as explained above. Of that list of 143 stars 
(and with the corrections described two paragraphs above), 
we exclude 31 stars that have calculated 
3D radial distances $>$15 pc from the cluster center. 
By plotting the remaining stars on a CMD we identify four interlopers that lie 
below the cluster sequence and one white dwarf. The white dwarf has $T_{eff}$ $\approx$ 10,200 K
and log $g$ $\approx$ 8.18 (Gentile Fusillo et al. 2019) which corresponds to a cooling age 
of $\sim$750 Myr (Dufour et al 2017), and therefore it cannot be a member of this $\sim$40 Myr old 
cluster (see third paragraph in Section 3).  
That leaves 107 {\it Gaia} sources, and with the recognition that the binary HD 21434 
(Mason et al 2015) is unresolved by {\it Gaia} (see Table 1), we present 108
cluster members and candidates (Tables 1$-$3) in a spherical volume of radius
15 pc.  The ranges in {\it Gaia} parameters
for this resulting set are:
\begin{center}
8.6 mas $<$ parallax $<$ 10.8 mas \\
29.3 mas/yr $<$ pmra $<$ 43.2 mas/yr \\
-9.7 mas/yr $<$ pmdec $<$ 0.9 mas/yr \\
\end{center}

Repeating the original search over a smaller region of R.A.-decl. angular distance =
4.4 degrees (8 pc), yields about 3/4 of the stars of the 15 pc search. 
In this inner 8 pc search region, relaxing the parallax or proper motion constraint 
boundaries adds 
very few (less than a handful) additional stars, and a similar exercise within the
even more compact region of the
tidal radius (3.5 pc, see Section \ref{sec:discussion}) finds just one or two
additional foreground sources.  This suggests a significant 
overabundance of stars within the range of the adopted constraints, as 
expected for a true cluster.  

Outside the 8 pc region, relaxing the search constraint boundaries can 
pick up a considerable number of additional stars (especially with a 
widened parallax range).
However, only a handful would actually make it to the $\chi$$^1$ For cluster candidate list,
as the vast majority of those extra stars do not pass our criteria of, (1) having
3D radial distance $<$15 pc from the cluster center, and (2) needing to be 
aligned with the cluster sequence on the CMD.
Thus, we consider our chosen {\it Gaia} search range to be an appropriate
set of limits for finding bona fide $\chi$$^1$ For members and candidates.

%\begin{table*}[h!]
\begin{table*}
\caption{$\chi$$^1$ For Cluster: Core Members}
\begin{center}
\begin{tabular}{lclccrrrlr}
\hline 
\hline
ChiFor & 2MASS 	&	SIMBAD  & $\Delta d$  & Spectral & Parallax  & $M_G$ &	 Mass & Comments \\
\#	&	(J2000)	&  Name	& (pc)\tablenotemark{a}	  & 	Type\tablenotemark{b} & (mas)  & (mag) & ($M_\sun$)\tablenotemark{c} & \\
\hline
1	&	03255188-3556254	&	CD-36 1289	&	0.23	&	G4	&	9.621	 (0.028)	&	4.65	&	1.00	&		\\
2A	&	03280743-3554434	&	CD-36 1309	&	0.40	&	F8	&	9.604	 (0.027)	&	4.13	&	1.20	&	Binary\tablenotemark{e}	\\
2B	&	03280941-3554365	&		&	0.41	&	M5	&	9.604	 (0.027)	&	11.50	&	0.20	&	Binary	\\
3	&	03265760-3555314	&		&	0.49	&	K8	&	9.556	 (0.024)	&	7.31	&	0.60	&		\\
4A	&	03281103-3551148	&	$\chi$$^3$ For	&	0.50	&	A1	&	9.625	 (0.039)	&	1.42	&	2.40	&	Binary	\\
4B	&	03281151-3551123	&	$\chi$$^3$ For B	&	0.50	&	G0	&	9.625	 (0.039)	&	5.06	&	1.10	&	Binary	\\
5	&	03255583-3555151	&	$\chi$$^1$ For	&	0.57	&	A1	&	9.564	 (0.035)	&	1.27	&	2.40	&		\\
6	&	03255277-3601161	&		&	0.91	&	M3.5	&	9.530	 (0.409)	&	9.49	&	0.30	&		\\
7 (A+B)	&	03255897-3557299	&	HD 21434	&	0.97	&	A9	&	   9.59 \tablenotemark{d}	 (0.04  \tablenotemark{d})	&	  \tablenotemark{d}    	&	3.4  \tablenotemark{d} 	&	Binary	\\
\hline
\end{tabular}
\end{center}
\tablenotetext{a}{$\Delta d$ is the distance of a star from the cluster center which is taken to be the average position of the A1-type stars $\chi$$^1$ For and $\chi$$^3$ For.  Based on four stars with published radial velocities ($\chi$$^1$ For itself, $\chi$$^3$ For, CD-36 1289, and ChiFor 3), the UVW of the core is -12.32$\pm$0.45, -21.46$\pm$0.67, -4.64$\pm$1.39 km s$^{-1}$ (see Section 4).}

\tablenotetext{b}{Spectral types given in Tables 1-3 are usually from SIMBAD when available (see Section 3), otherwise from photometric colors as, for example, those given in Table 4.}

\tablenotetext{c}{Stellar masses are an average of those given in Table 15.8 of the 4th edition of
Allen's Astrophysical Quantities and
http://www.pas.rochester.edu/$\sim$emamajek/EEM\_dwarf\_UBVIJHK\_colors\_Teff.txt.}

%\tablenotetext{c}{Stellar masses are an average of those given in Table 15.8 of the 4th edition of Allen's 
%Astrophysical Quantities and}
%\footnote{ http://www.pas.rochester.edu/$\sim$emamajek/EEM_dwarf_UBVIJHK_colors_Teff.txt}}

\tablenotetext{d}{HD 21434 is an equal mass (each star is $\sim$1.7 $M_\sun$) 0.2'' separation binary (Mason et al 2015).  Its {\it Gaia} DR2 parallax value of 9.08$\pm$0.17 mas is very likely corrupted by the binarity (note the relatively large quoted error for such a bright star).  We assume a parallax for the binary equal to the mean parallax of the other core members, i.e. 9.59$\pm$0.04 mas.  Then the $M_G$ of each component of the HD 21434 binary is 2.89 mag.}

\tablenotetext{e}{See Note added in proof for radial velocity and other data for ChiFor 2A.}
\label{tab:core}
\end{table*}

%\begin{table*}[h!]
\begin{table*}
\caption{$\chi$$^1$ For Cluster: Likely Members between the Core and the Tidal Radius\tablenotemark{a}}
\begin{center}
\begin{tabular}{lclccrrrlr}
\hline 
\hline
ChiFor & 2MASS 	&	SIMBAD  & $\Delta d$  & Spectral & parallax  & $M_G$ &	 Mass & Comments	\\
\#	&	(J2000)	& Name	&  (pc)	  & 	Type & (mas)  & (mag)	& ($M_\sun$)	&  \\
\hline

8	&	03283738-3539286	&		&	1.22	&	M4	&	9.692	 (0.176)	&	10.60	&	0.20	&		\\
9	&	03290633-3541070	&		&	1.39	&	M2	&	9.499	 (0.027)	&	8.60	&	0.40	&		\\
10	&	03261876-3637064	&		&	1.54	&	M5.5	&	9.667	 (0.132)	&	12.53	&	0.15	&		\\
11	&	03321206-3539433	&		&	1.99	&	M6	&	9.560	 (0.113)	&	12.33	&	0.15	&		\\
12	&	03251321-3515552	&		&	2.02	&	M4.5	&	9.743	 (0.078)	&	11.60	&	0.20	&		\\
13	&	03285448-3650168	&		&	2.16	&	M4	&	9.702	 (0.063)	&	10.79	&	0.20	&		\\
14	&	03270606-3445494	&		&	2.36	&	M3	&	9.493	 (0.037)	&	9.40	&	0.35	&	IR excess	\\
15A	&	03251011-3445277	&		&	2.39	&	M2	&	9.508	 (0.029)	&	8.61	&	0.40	&	Binary; See table notes	\\
15B	&	03251033-3445206	&		&	2.40	&	M2	&	9.508	 (0.031)	&	8.83	&	0.40	&	Binary; IR excess	\\
16	&	03292421-3514293	&		&	2.68	&	M5	&	9.812	 (0.072)	&	11.40	&	0.20	&	IR excess	\\
17A	&	03303532-3512133	&		&	2.81	&	M3.5	&	9.405	 (0.051)	&	10.06	&	0.30	&	Binary	\\
17B	&	03303488-3512335	&		&	2.81	&	M3.5	&	9.405	 (0.051)	&	10.08	&	0.30	&	Binary	\\
18	&	03343867-3529039	&		&	2.90	&	K5	&	9.611	 (0.022)	&	6.54	&	0.70	&	UV variable	\\
19	&	03243659-3508586	&		&	2.93	&	M4	&	9.831	 (0.057)	&	10.28	&	0.20	&		\\
20	&	03285610-3617125	&		&	3.03	&	M4.5	&	9.344	 (0.071)	&	11.12	&	0.20	&		\\
21	&	03221999-3638138	&	CD-37 1263	&	3.07	&	G5	&	9.801	 (0.026)	&	4.82	&	1.00	&	RV=14.31$\pm$2.57	\\
22	&	03293515-3540429	&		&	3.17	&	M5.5	&	9.330	 (0.136)	&	12.74	&	0.15	&		\\
23	&	03310150-3713188	&		&	3.26	&	M4	&	9.453	 (0.058)	&	10.27	&	0.20	&		\\
24	&	03343163-3501033	&		&	3.27	&	M2	&	9.532	 (0.081)	&	8.73	&	0.40	&	IR excess	\\
25A	&	03185295-3607375	&	HD 20707	&	3.39	&	F3	&	9.463	 (0.030)	&	2.98	&	1.50	&	Binary; IR excess	\\
25B	&	03185312-3607489	&	HD 20707B	&	3.39	&	M4	&	9.463	 (0.030)	&	10.20	&	0.20	&	Binary; IR excess	\\
26A	&	03231580-3435524	&		&	3.56	&	M3	&	9.394	 (0.043)	&	9.65	&	0.35	&	Binary	\\
26B	&	03231705-3435472	&		&	3.56	&	M8	&	9.394	 (0.043)	&	14.53	&	0.10	&	Binary	\\
27	&	03351859-3652183	&		&	3.80	&	M3.5	&	9.735	 (0.055)	&	9.96	&	0.30	&		\\
\hline
\end{tabular}
\end{center}

\tablenotetext{a}{Tables 1 and 2 together list the likely members of the $\chi$$^1$ For cluster contained within its tidal radius ($\sim$3.5 pc, see Section 4).  ChiFor 27 is included in this table because, within the uncertainties, it might be within the tidal radius.  See Section 3 for a discussion of the excess infrared emission of ChiFor 15B.   For CD-37 1263, RV is its radial velocity in km s$^{-1}$.} 
\label{tab:tidal}
\end{table*}

\section{Results}

\begin{figure}
\plotone{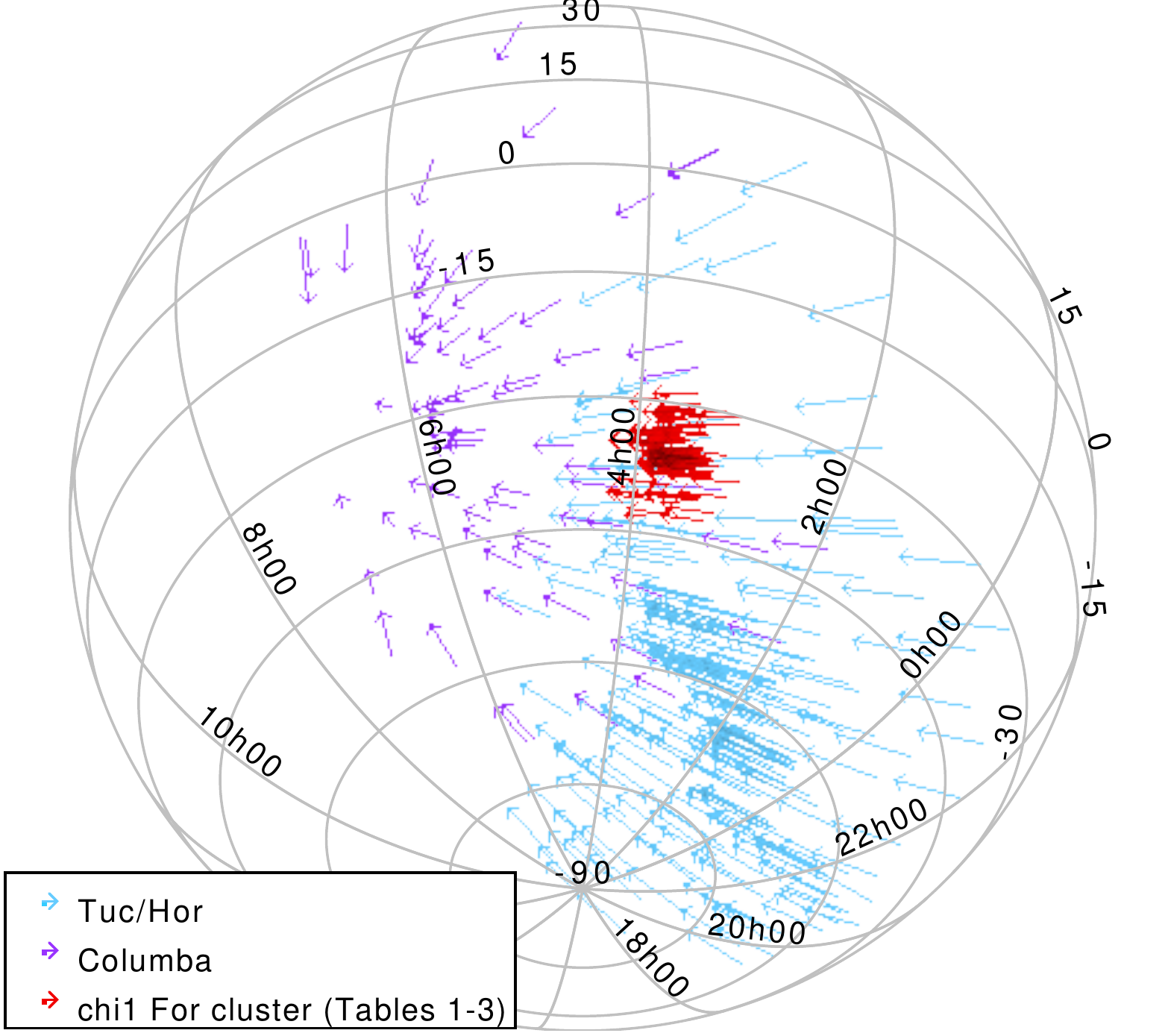}
\caption{Slant orthographic projection in which the arrows indicate the proper motion in the plane of the sky of proposed $\chi$$^1$ For member stars (Tables 1$-$3) and for stars in the two Associations, Tuc-Hor and Columba. Data are from Lee \& Song (2019) and the {\it Gaia} DR2 catalog. As expected for comoving stars, the sky-plane motions of all three groups converge on a common sky position (near 6h45m, -30d).}
\label{fig:pmplot}
\end{figure}

Stars considered as potential members of the $\chi$$^1$ For cluster appear 
in Tables 1-3 and in Figures \ref{fig:skyplot}-\ref{fig:pmplot}.  For stars with SIMBAD resolvable names, the listed spectral types are usually from SIMBAD, but there are a few exceptions. For example, SIMBAD does not give a spectral type for $\chi$$^3$ For B.  For the other late-F and G-type stars (other than the close binary HD 21955) we checked the SIMBAD spectral types versus $M_G$, $B-V$ from Tycho, and $V-Ks$.    We then estimated photometric spectral types with the aid of the colors given at the http: URL in footnote ``c'' of Table 1. Then, for CD-36 1309, CD-37 1263, and CD-37 1224 we adopted the SIMBAD spectral type.  But for CD-36 1289 we adopt G4 (rather than F8) and for HD 24121 we adopt G3 (rather than F8/G0). 

The listed spectral types of M- and K-type
stars were deduced by comparison with stars in the $\sim$40 Myr old Tuc-Hor (Kraus et al 
2014) Association.  Specifically, we constructed Table 4 from stars in Table 2 of Kraus et al (2014) and used a combination of $M_G$, $M_{Ks}$, $G-Ks$, and $Ks-W2$ to estimate the M-star spectral types listed in Tables 1-3.  But for stars with excess infrared emission, $Ks-W2$ was not used. Thus, the ``spectral types'' given in Tables 1-3 for K- and M-type stars are based on photometry, not spectroscopy, and are probably good to $\sim$1/2 subclass.

Figure \ref{fig:cmd} includes field stars within 25 pc of Earth; at a given color these stars
plot below the $\chi$$^1$ For cluster stars and Columba Association stars. Stars in $\chi$$^1$ For 
and in Columba are thus younger than the field stars.  Figure \ref{fig:cmd} shows that the stars plotted in Figures  \ref{fig:skyplot} and \ref{fig:cmd} have ages that are similar to Columba stars, that is, $\sim$40 Myr 
(Mamajek 2016; Lee \& Song 2019).  Stars that lie just above (within 0.75 mag of) the main locus are most likely binaries unresolved by {\it Gaia}; taken together out to the tidal radius (see Section 4), the companion stars likely add about one solar mass of material to the cluster.   Although not shown here, stars in Tuc-Hor (Kraus et al. 2014; Lee \& Song 2019) would also plot along the same locus. An age for the $\chi$$^1$ For cluster between 500 and 600 Myr is given in papers by P\"ohnl \& Paunzen (2010), 
Kharchenko et al.~(2016), and Yen et al.~(2018), but Figure \ref{fig:cmd} 
demonstrates that the cluster is of the order of 10 times younger.  We note that Mamajek (2016) suggested a cluster age of $\sim$30 Myr
based on a few previously identified cluster stars with saturated X-ray emission; the number of stars in that sample was much less than the number of cluster candidates we consider here that were detected in X-rays (see the last paragraph of Section 3).

Radial velocities are available for only a few stars in Tables 1-3; see note ``a'' in Table 1 and
the comments column in Tables 2 \& 3. Section 4 below includes a discussion of the cluster
membership likelihood of each star with a currently known radial velocity.   Lacking a known radial velocity, likely members are chosen based on parallax, proper motion, and location on 
a {\it Gaia} CMD (Figure \ref{fig:cmd}).  For potential members with known 
radial velocities, UVW can be calculated and compared with the space motions 
of members of the Columba and Tuc-Hor associations; the agreement is usually 
excellent (see Section 4).

Table 3 presents stars that lie outside of the tidal radius (see Section 4) 
of the $\chi$$^1$ For cluster. Each of these stars could be  
an escaped member of the cluster or a member of either the Columba or 
Tuc-Hor Associations (Lee \& Song 2019).  Based on the relative frequency of dusty debris
disks at $\chi$$^1$ For cluster stars relative to members of Tuc-Hor (see discussion in Section 4), 
many Table 3 stars beyond 8 pc from the cluster center probably are 
associated with Columba or Tuc-Hor rather than with
$\chi$$^1$ For.  As has been realized for a decade 
(Torres et al 2008), Columba stars are similar to stars in Tuc-Hor 
(Zuckerman et al 2001), both in age and space motions through the Milky Way. 
However, published studies of these associations (Kraus et al 2014; Lee \& Song 2019) 
are based on catalogs and 
spectral measurements that are largely insensitive to stars as far away as 
the $\chi$$^1$ For cluster.  
Clarification of whether or not Columba and Tuc-Hor members are to be found 
100 pc from Earth, and even beyond, remains as a task for the future.  
Together, the Columba and Tuc-Hor Associations and the $\chi$$^1$ For 
cluster constitute a comoving and coeval stellar cohort that 
spans about half of the southern sky between R.A. 20h and 8h and 
south of decl. -15 deg (Figure \ref{fig:pmplot}).

\begin{figure*}
\plotone{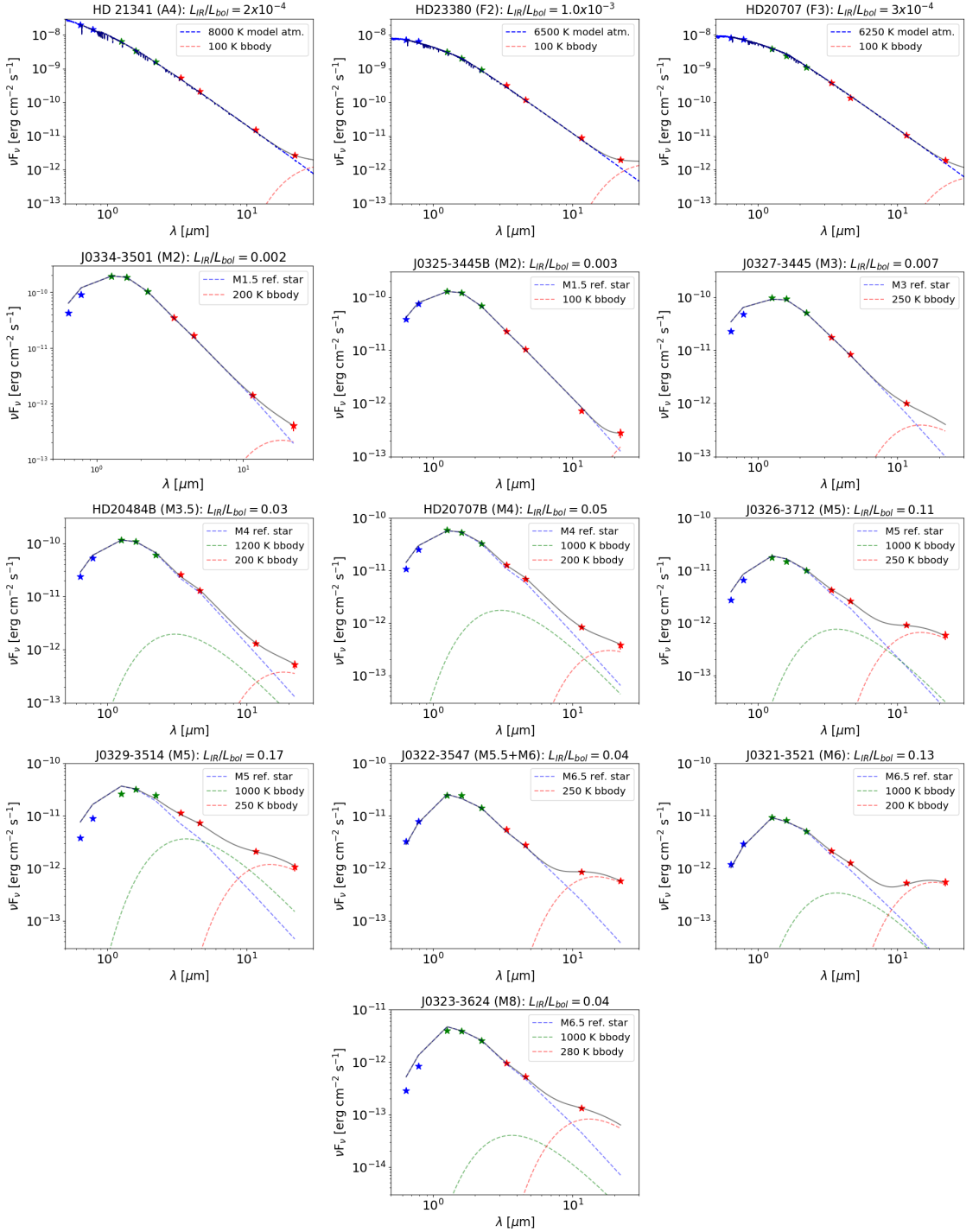}
\caption{$\chi$$^1$ For cluster stars for which we find evidence of
  excess infrared emission above the stellar photosphere. The blue, green,
  and red symbols indicate {\it Gaia} ($G$ and $G_{RP}$ band), 2MASS ($J$, $H$, $K$
  band) fluxes, and {\it WISE} ($W$1, $W$2, $W$3, $W$4 band) fluxes,
  respectively. The error bars on the fluxes are smaller than the symbol sizes
  in almost all cases. The measured SEDs for the M-type stars are overlaid with a model consisting of a stellar photosphere plus a one- or two-component blackbody, where these blackbodies represent the excess emission due to circumstellar dust. The models overlaid on the SEDs of the A- and F-type stars (HD 21341, HD 23380, and HD 20707) consist of stellar model atmospheres obtained from the Bosz library maintained by STScI (see {\tt https://archive.stsci.edu/prepds/bosz/})
and blackbodies. The resulting estimates of the fractional IR luminosities and inferred blackbody temperatures are indicated for each star.  The SED of J0325-3445B (= ChiFor 15B) has been modified in a way discussed
in Section 3 of the text.}
\label{fig:sed}
\end{figure*}

Figure \ref{fig:sed} presents the spectral energy distributions (SEDs) of the 13 stars 
listed in Tables 1-3 for which we find evidence in the {\it WISE} catalog (Wright 
et al 2010) of infrared emission in excess of that from the stellar 
photosphere, overlaid with models consisting of photosphere plus a one- or 
two-component blackbody. The spectral type of a given star was used to 
select appropriate photospheric reference stars, where these reference stars 
were drawn from Table 3 in Pecaut \& Mamajek (2013), as well as from our 
list of $\chi$$^1$ For cluster candidates showing no evidence of IR 
excess. All of the reference stars have colors (from {\it Gaia}-G through {\it WISE}-4 
bands) that agree, within $\sim$0.1 mag, with the spectral type-dependent colors 
listed in Table 5 of Pecaut \& Mamajek. For the A4, F3, and F2 stars (HD 21341, 
HD 20707, and HD 23380 respectively), the stellar photospheres are 
represented by model atmospheres with appropriate effective temperatures for 
these spectral types. To establish the IR excess for each star, the 
reference star (or model atmosphere) SEDs have been rescaled to match a 
star's photospheric emission at the $H$ or $K$ band. For each star we obtained, by 
eye, a suitable match of one- or two-component blackbody emission (plus 
underlying photosphere) to the star's $W$1 through $W$4 fluxes. The resulting 
blackbody ``fits'' to the long-wavelength portions of the SEDs in Figure \ref{fig:sed}
should not be regarded as unique; rather, they provide estimates of the 
fractional infrared luminosities of these stars, as well as rough 
approximations of the circumstellar dust temperatures.

For most stars, the one- or two-component blackbody models provide good 
matches to the observed SEDs.  For a few 
M-type stars, the match to reference star photospheric (especially $G$ and $G_ 
{RP}$ band) fluxes is not optimal. Some of these stars may be affected by 
reddening due to interstellar or circumstellar dust. In these cases we opted 
for the coolest reference star that matched the near-IR fluxes, to obtain 
the most conservative estimates of IR excess. 

Two stars deserve special mention.   The SED of the star ChiFor 15B listed in Table 2 and shown in 
Figure \ref{fig:sed} indicates excess emission in only the {\it WISE} $W$4 band.  But if the {\it Gaia}, 2MASS 
and {\it WISE} magnitudes are taken at face value, then an SED plot would indicate excess emission in
all four {\it WISE} bands.   ChiFor 15A and B appear as separate stars in the {\it Gaia} and 2MASS catalogs and should also appear in {\it WISE} as two separate stars.  But only one star appears.  Evidently, the separation of the two stars ($\sim$7'') is such that there was confusion in separately assigning {\it WISE} Point Source
Catalog fluxes to the components of the ChiFor 15AB binary system.  Therefore, to prepare the SED of ChiFor 15B in the panel shown in Figure \ref{fig:sed}, we have added 0.75 magnitudes to the magnitudes listed at all four {\it WISE} bands. 

The Table 3 star J0338-3334 has apparent excess emission in the $W$3 and $W$4 bands.  But there is a ``dwarf galaxy'' 8'' away from the star.  It is thus not possible, based on existing information, to decide if 
the apparent excess is from dust associated with the star or from contamination by the background 
galaxy, or both.

\startlongtable
\begin{deluxetable*}{clccrrrl}
%\begin{center}
\tablecaption{$\chi$$^1$ For Cluster: Potential Members Outside the Tidal Radius} \label{tab:outside}
%\begin{tabular}
%\hline
%\hline
\tablehead{
\colhead{2MASS} & \colhead{SIMBAD}  & \colhead{$\Delta d$} &  \colhead{Spectral}  & \colhead{parallax}  & \colhead{$M_G$} & \colhead{Mass} & \colhead{Comments}  \\
\colhead{(J2000)} & \colhead{Name} & \colhead{(pc)} &  \colhead{Type} &  \colhead{(mas)} & \colhead{(mag)} & \colhead{($M_\sun$)} & \colhead{}
}

%\colnumbers
\startdata
03163272-3541282	&	HD 20484	&	4.05	&	A3	&	9.705	 (0.036)	&	1.94	&	2.20	&	Binary	\\
03163154-3541337	&	HD 20484B	&	4.06	&	M3.5	&	9.705	 (0.036)	&	9.39	&	0.30	&	Binary; IR excess	\\
03353527-3702540	&		&	4.28	&	M3.5	&	9.792	 (0.050)	&	9.88	&	0.30	&		\\
03190162-3617152	&		&	4.30	&	M5.5	&	9.328	 (0.111)	&	12.18	&	0.15	&		\\
03251281-3709096	&	HD 21341	&	4.33	&	A4	&	9.944	 (0.029)	&	2.13	&	2.00	&	Binary; IR excess	\\
	&	HD 21341B	&	4.33	&	M3.5	&	9.944	 (0.029)	&	9.85	&	0.30	&	Binary	\\
03221529-3547192	&		&	4.40	&	M5.5	&	9.985	 (0.181)	&	12.18	&	0.15	&	Binary; IR excess	\\
B	&		&	4.40	&	M6	&	9.986	 (0.182)	&	12.62	&	0.15	&	Binary; see table notes	\\
03292874-3604552	&		&	4.53	&	M5.5	&	9.208	 (0.106)	&	12.29	&	0.15	&		\\
03210190-3521218	&		&	4.88	&	M6	&	10.007	 (0.109)	&	12.72	&	0.15	&	IR excess	\\
03265980-3712176	&		&	5.16	&	M5	&	9.197	 (0.089)	&	11.64	&	0.20	&	IR excess	\\
03261930-3744352	&		&	5.22	&	M3	&	9.981	 (0.037)	&	9.32	&	0.35	&		\\
03184285-3538356	&		&	5.58	&	M4	&	9.191	 (0.076)	&	11.07	&	0.20	&		\\
03132905-3609026	&		&	5.61	&	M3.5	&	9.839	 (0.057)	&	9.78	&	0.30	&		\\
03151711-3702301	&	HD 20379	&	5.94	&	F5	&	9.935	 (0.026)	&	3.52	&	1.40	&	RV=17.59$\pm$0.97	\\
03121329-3706151	&		&	6.15	&	M1.5	&	9.778	 (0.390)	&	8.49	&	0.45	&		\\
03214475-3309494	&		&	6.19	&	M3	&	9.897	 (0.051)	&	9.66	&	0.35	&	Binary	\\
B	&		&	6.19	&	M3.5	&	9.897	 (0.051)	&	9.93	&	0.30	&	Binary	\\
03252401-3706057	&		&	6.26	&	M3.5	&	10.169	 (0.046)	&	9.83	&	0.30	&		\\
03380114-3334215	&		&	6.44	&	M5.5	&	9.362	 (0.174)	&	12.26	&	0.15	&	See table notes	\\
03243582-3221528	&		&	6.58	&	M3	&	9.490	 (0.091)	&	9.31	&	0.35	&		\\
03320810-3413201	&		&	6.60	&	M3.5	&	9.114	 (0.048)	&	9.71	&	0.30	&		\\
03353519-3304505	&		&	6.67	&	K7	&	9.342	 (0.026)	&	7.62	&	0.60	&	RV=20.00$\pm$2.04	\\
03230624-3624131	&		&	6.87	&	M8	&	9.024	 (0.331)	&	14.04	&	0.10	&	IR excess	\\
03274299-3214077	&		&	6.90	&	K6	&	9.430	 (0.027)	&	7.08	&	0.70	&	RV=18.81$\pm$2.23	\\
03160305-3725222	&	CD-37 1224	&	7.05	&	G8	&	10.091	 (0.023)	&	4.83	&	0.90	& See table notes \\
03273888-3755432	&		&	7.10	&	M6	&	9.074	 (0.134)	&	12.89	&	0.15	&		\\
03410202-3251005	&		&	7.65	&	M4	&	9.671	 (0.072)	&	10.82	&	0.20	&		\\
03432786-3330081	&	HD 23380	&	7.82	&	F2	&	9.401	 (0.033)	&	3.15	&	1.50	&	IR excess	\\
03384658-3337289	&		&	8.08	&	M4	&	9.126	 (0.053)	&	10.09	&	0.20	&		\\
03262402-4008431	&		&	8.24	&	M6	&	9.867	 (0.106)	&	12.76	&	0.15	&		\\
03231537-3247371	&		&	9.02	&	M6	&	9.003	 (0.160)	&	12.87	&	0.15	&		\\
03234843-3652388	&		&	9.20	&	M4	&	10.502	 (0.061)	&	10.20	&	0.20	&		\\
03364806-3239493	&		&	9.45	&	M3.5	&	9.040	 (0.045)	&	9.95	&	0.30	&		\\
03160047-3729058	&		&	9.52	&	M4	&	8.905	 (0.083)	&	10.87	&	0.20	&		\\
03465367-3916341	&		&	9.82	&	M3	&	9.355	 (0.042)	&	9.59	&	0.35	&		\\
03341952-3431171	&		&	10.11	&	M4	&	8.803	 (0.081)	&	11.01	&	0.20	&		\\
02592723-3713404	&		&	10.37	&	M5.5	&	9.580	 (0.113)	&	12.24	&	0.15	&		\\
03310546-3039084	&		&	10.40	&	M4	&	9.253	 (0.078)	&	10.58	&	0.20	&		\\
03424074-3439560	&		&	10.66	&	M0	&	8.863	 (0.022)	&	8.07	&	0.55	&	Binary	\\
03424123-3439413	&		&	10.66	&	M4	&	8.863	 (0.022)	&	10.34	&	0.20	&	Binary	\\
03313001-3358142	&		&	10.78	&	K5	&	8.754	 (0.024)	&	6.41	&	0.70	&	RV=22.0$\pm$0.5	\\
03412520-3258427	&		&	11.38	&	M6	&	8.874	 (0.135)	&	12.57	&	0.15	&		\\
03304324-3034477	&		&	11.43	&	M3	&	9.079	 (0.047)	&	9.39	&	0.35	&		\\
03142554-3007004	&		&	11.54	&	M2.5	&	9.592	 (0.036)	&	8.93	&	0.40	&	RV=9.6$\pm$8.8	\\
03200945-4017402	&	HD 20854	&	11.62	&	F3	&	8.912	 (0.026)	&	3.70	&	1.50	&	RV=11.9$\pm$0.5	\\
03041564-3919129	&		&	11.65	&	M4.5	&	10.127	 (0.060)	&	11.17	&	0.20	&		\\
03055457-3111275	&		&	11.71	&	M5	&	9.642	 (0.086)	&	11.25	&	0.20	&		\\
03240056-3055429	&		&	11.80	&	K9	&	8.953	 (0.031)	&	7.73	&	0.55	&	RV=16.69$\pm$3.24	\\
03533894-3443562	&	i Eri	&	11.86	&	B6.5	&	9.057	 (0.164)	&	-0.16	&	5.00	&	See table notes	\\
03535952-3442463	&		&	11.96	&	M4.5	&	9.059	 (0.070)	&	11.35	&	0.20	&	See table notes	\\
03432006-3437384	&		&	12.15	&	M5	&	8.737	 (0.099)	&	11.68	&	0.20	&		\\
03312080-3030588	&	HD 21955	&	12.21	&	G7	&	8.983	 (0.102)	&	4.51	&	1.60	&	Binary; see table notes	\\
03473987-4114014	&		&	12.34	&	M5	&	9.412	 (0.081)	&	11.52	&	0.20	&		\\
03390055-3928559	&		&	12.41	&	M4	&	8.787	 (0.061)	&	10.26	&	0.20	&		\\
03212391-4033418	&		&	12.52	&	M8	&	8.839	 (0.518)	&	14.77	&	0.10	&		\\
03243747-3041230	&		&	12.54	&	M2	&	8.899	 (0.038)	&	8.34	&	0.40	&	Binary	\\
03243702-3041213	&		&	12.55	&	M4.5	&	8.899	 (0.038)	&	10.75	&	0.20	&	Binary	\\
03533447-3445140	&		&	12.67	&	M3.5	&	8.937	 (0.044)	&	9.89	&	0.30	&	See table notes	\\
03303548-3020058	&	HD 21881	&	12.78	&	F3	&	8.937	 (0.038)	&	2.90	&	1.50	&	RV=30.48$\pm$6.22	\\
03485139-4038356	&	HD 24121	&	12.86	&	G3	&	10.136	 (0.020)	&	4.65	&	1.00	&	RV=21.06$\pm$0.53	\\
03355621-4245510	&		&	13.01	&	M3.5	&	9.766	 (0.059)	&	10.03	&	0.30	&		\\
03105227-4139227	&		&	13.07	&	M3	&	10.114	 (0.039)	&	9.52	&	0.35	&		\\
03363743-4231544	&		&	13.19	&	K7	&	9.238	 (0.021)	&	7.43	&	0.60	& 	RV=30.68$\pm$1.84	\\
03503915-4056499	&		&	13.24	&	M5	&	9.203	 (0.078)	&	11.64	&	0.20	&		\\
03290832-2947065	&		&	13.34	&	M3	&	8.966	 (0.041)	&	9.34	&	0.35	&		\\
03103637-3502240	&		&	13.46	&	M0	&	8.615	 (0.015)	&	8.00	&	0.55	&	RV=18.1$\pm$2.0	\\
03443839-2929157	&		&	13.96	&	M2	&	9.263	 (0.029)	&	8.59	&	0.40	&		\\
03311122-2832184	&		&	13.98	&	M0	&	9.963	 (0.038)	&	8.02	&	0.55	&	RV=23.05$\pm$1.33	\\
03060195-3018206	&		&	14.31	&	M4.5	&	9.064	 (0.082)	&	11.41	&	0.20	&		\\
02551799-3344443	&		&	14.73	&	M3.5	&	10.381	 (0.044)	&	9.73	&	0.30	&		\\
03245822-4039380	&		&	14.79	&	M3.5	&	10.844	 (0.050)	&	9.90	&	0.30	&		\\
03150651-4237352	&		&	14.83	&	M4	&	8.978	 (0.187)	&	10.67	&	0.20	&		\\
03370343-3042318	&		&	15.05	&	M5.5	&	8.675	 (0.090)	&	11.65	&	0.15	&	RV=11.9$\pm$3.5 	\\
\enddata
\tablecomments{ 2MASSJ03221529-3547192 is a close binary, unresolved by 2MASS and {\it WISE}.  We designate the secondary with the letter B. Thus it is not possible to tell if the
excess emission in the {\it WISE} $W$3 and $W$4 bands is from one or both of the
stars.  HD 21955 is a close G7 + K5 binary.  J0338-3334 may have excess infrared emission (see discussion in Section 3).  J0353-3445 and J0353-3442 may be distant
(10,000 AU and 29,000 AU separation) companions of i Eri.  RV is radial velocity in km s$^{-1}$.  See Note added in proof for radial velocity and other data for CD-37 1224.}
\label{tab:outside}
\end{deluxetable*}

\begin{table}
\caption{Tucana-Horologium: M-type Star Magnitudes and Colors }
\begin{center}
\begin{tabular}{lccccc}
\hline 
\hline
Spec & $M_G$ & $M_{Ks}$  & $G-Ks$  & $Ks-W2$ & $Ks-W2$  	\\
Type\tablenotemark{a} & (mag)  & (mag) &  (mag) &  (mag) &  (P\&M)  \\
\hline
M2	&	8.78	& 5.25 &	3.53 &	0.22	&	0.22  \\
M3	&	9.39	& 5.64 &	3.75	&  0.26	&	0.28  \\
M3.5	 &	9.76	&  5.92 &	3.84	&  0.33 &.   \\
M4	&	10.69 & 6.67 &	4.02	&  0.37	&	0.35  \\
M4.5	 &	11.61 & 	7.39	&  4.22 &	0.41 &  \\
M5	&	11.77 &  	7.44	&  4.33 &	0.42	&	0.41 \\
M5.5	 &	12.24 &	7.73	&  4.51 &	0.46	&  \\
\hline
\end{tabular}
\end{center}
\tablenotetext{a}{Each entry in the table is based on three stars in Table 2 of Kraus et al  (2014) except M5.5 (two stars).  P\&M refers to colors given in Pecaut \& Mamajek (2013).}
%\label{tab:core}
\end{table}

%\begin{table*}[h!]
\begin{table*}
\caption{$\chi$$^1$ For Cluster Members and Candidates with UV and/or X-Ray Counterparts}
\begin{center}
\begin{tabular}{llclllr}
\hline 
\hline
ChiFor & Designation	 &  $G$  &	NUV & 	NUV-G	&	$F_x$ (source)	&	 log($L_x$/$L_{bol}$)	\\
\#      &	&	(mag)	&	(mag)	&	(mag)	&	(10$^{-12}$ erg/cm$^2$/s)  \\		
\hline	
1 & CD-36 1289 	&	9.73	&	\ldots	& \ldots	&	5.0 ({\it XMM})	&	-2.86	\\
 &	&		&		&		&	1.1 ({\it ROSAT})	&	-3.53	\\
2A & CD-36 1309\tablenotemark{a}	&	9.22	&	\ldots	&	\ldots	&	1.7\tablenotemark{a} ({\it ROSAT})	&	-3.54\tablenotemark{a}\\
14 &03270606-3445494 	&	14.51	&	21.73 (0.26)	&	7.22	&	\ldots	&	\ldots	\\
15A & 03251011-3445277\tablenotemark{b} 	&	13.72 &	20.70\tablenotemark{b} (0.17)	&	7.73\tablenotemark{b}	& \ldots	& \ldots	\\
15B & 03251033-3445206\tablenotemark{b}	&	13.94 &	20.70\tablenotemark{b} (0.17)	&	7.51\tablenotemark{b}	& \ldots	& \ldots	\\
17A &03303532-3512133 	&	15.19	&	23.17 (0.43)	&	7.98	&	\ldots	&	\ldots	\\
% not in 2017 AIS: 18B & 03303488-3512335 	&	15.22	&	22.88 (0.46)	&	7.66	&	\ldots	&	\ldots	\\
18 &03343867-3529039 	&	11.63	&	19.66 (0.10)	&	8.03	&	0.29 ({\it ROSAT})	&	-3.6	\\
21 & CD-37 1263 	&	9.87	&	15.23 (0.01)	&	5.36	&	2.7 ({\it XMM})	&	-3.12	\\
24 &03343163-3501033	&	13.83	&	20.80 (0.21)	&	6.97	&	\ldots	&	\ldots	\\
25A & HD 20707 	&	8.10	&	12.397 (0.002)	&	4.30	&	\ldots	&	\ldots	\\
26A &03231580-3435524	&	14.79	&	22.03 (0.39)	&	7.24	&	\ldots	&	\ldots	\\
\ldots & HD 20484 	&	7.01	&	11.738 (0.002)	&	4.74	&	\ldots	&	\ldots	\\
\ldots & HD 20484B\tablenotemark{a}	&	14.45	&	\ldots	&	\ldots	&	0.44\tablenotemark{a} ({\it ROSAT})	&	-3.16\tablenotemark{a}\\
\ldots & 03353527-3702540	& 14.93	&	21.74 (0.29)	&	6.81 &	\ldots	&	\ldots \\
\ldots & HD 21341B\tablenotemark{a} & 14.86 & \ldots & \ldots & 0.06\tablenotemark{a} ({\it ROSAT})	& -3.99\tablenotemark{a}\\
\ldots & 03265980-3712176	& 16.82	&	22.30 (0.38)	&	5.48 &	\ldots	&	\ldots \\
\ldots & 03261930-3744352 	&	14.33	&	21.25 (0.20)	&	6.92	&	\ldots	&	\ldots	\\
\ldots & HD 20379	& 8.53	&	13.29 (0.01)	&	4.75  &	\ldots	&	\ldots	\\
\ldots & 03214475-3309494\tablenotemark{b}	&	14.68	&	21.52\tablenotemark{b} (0.22)	&	7.59\tablenotemark{b} & \ldots	&	\ldots	\\
\ldots & 03214475-3309494B\tablenotemark{b}&	14.95	&	21.52\tablenotemark{b} (0.22)	&	7.27\tablenotemark{b} & \ldots	&	\ldots	\\
\ldots & 03252401-3706057	&	14.79	&	20.95 (0.17)	&	6.16 & \ldots	&	\ldots	\\
\ldots & 03243582-3221528	&      14.42	&	21.48 (0.21)	&	7.05 & \ldots	&	\ldots	\\
\ldots & 03320810-3413201 	&	14.91	&	22.01 (0.35)	&	7.10	&	\ldots	&	\ldots	\\

\ldots & 03353519-3304505	&	12.77	&	20.33 (0.08)	&	7.56 & \ldots & \ldots \\
\ldots & 03274299-3214077	&	12.21	&	20.35 (0.11)	&	8.14	& \ldots & \ldots \\		
\ldots & CD-37 1224              &      9.81	&	15.47 (0.01)	&	5.66	& 0.88 ({\it ROSAT}) &	-3.54 \\
\ldots & 03465367-3916341	&	14.73	&	21.83 (0.48)	&	7.09 & \dots & \ldots \\
\ldots & 03424074-3439560\tablenotemark{a}	&	13.33	& 21.14 (0.28)	&	7.80	& 0.57\tablenotemark{a} ({\it ROSAT}) &	-2.89\tablenotemark{a} \\
\ldots & 03313001-3358142	&	11.70	&	19.47 (0.06)	&	7.78 & \ldots & \ldots \\
\ldots & 03304324-3034477	&	14.60	&	21.72 (0.26)	&	7.12 & \ldots & \ldots \\
\ldots  & HD 20854	&	8.95	&	13.27 (0.01)	&	4.32 & \ldots & \ldots \\
\ldots & 03240056-3055429	&	12.97	&	20.65	0.15	&	7.68 & \ldots & \ldots \\
\ldots  & HD 21955        &	9.74	&	14.75 (0.01)	&	5.01	& 1.6 ({\it ROSAT}) &	-3.39 \\
\ldots & 03390055-3928559	&		15.54	&	22.28 (0.25)	&	6.74 & \ldots & \ldots \\
\ldots & 03243747-3041230	&	13.59	&	21.12 (0.19)	&	7.52 & \ldots & \ldots \\
\ldots  & HD 21881	&	8.14	&	12.43 (0.01)	&	4.29 & \ldots & \ldots \\
\ldots  & HD 24121	&	9.62	&	14.97 (0.01)	&	5.35 & \ldots & \ldots \\
\ldots & 03355621-4245510	&		15.08	&	21.74 (0.36)	&	6.66			 & \ldots & \ldots \\
\ldots & 03105227-4139227	&		14.50	&	22.14 (0.43)	&	7.64	& 0.16 ({\it ROSAT})&	-3.12 \\
\ldots & 03363743-4231544	&	12.60	&	22.27 (0.45)	&	9.67			 & \ldots & \ldots \\
\ldots & 03290832-2947065	&	14.58	&	21.93 (0.27)	&	7.36 & \ldots & \ldots \\
\ldots & 03103637-3502240	&	13.32	&	21.50 (0.23)	&	8.18 & \ldots & \ldots \\
\ldots & 03311122-2832184	&	13.03	&	22.31 (0.38)	&	9.28 & \ldots & \ldots \\
\ldots & 03245822-4039380	&		14.72	&	22.46 (0.42)	&	7.74 & \ldots & \ldots \\
\hline
\end{tabular}
\end{center}
\tablenotetext{a}{These stars are components of visual binaries whose separations are
smaller than the {\it ROSAT} PSF. The {\it ROSAT} source flux is attributed to the primary or secondary based on the inferred values of  log($L_x$/$L_{bol}$).}
\tablenotetext{b}{{\it GALEX} UV source is roughly equidistant from components A and B of these binaries; listed NUV-G colors assume both components contribute equally to the UV flux.}
\label{tab:UV_Xray}
\end{table*}

Table 5 lists all UV photometry (from {\it GALEX}; Bianchi et al. 2017) and X-ray data (from the {\it ROSAT} 
All-Sky Survey and {\it XMM-Newton} Serendipitous Source Catalog (SSC)) available in the 
NASA/HEASARC archives\footnote{\url{https://heasarc.gsfc.nasa.gov/cgi-bin/W3Browse/w3browse.pl}}
for the cluster candidate members listed in Tables 1$-$3.  Search radii of 3'', 30'', and 5'' were 
used to identify {\it GALEX}, {\it ROSAT}, and {\it XMM} counterparts to cluster candidates. For 
{\it ROSAT} sources, the X-ray fluxes were obtained by applying a standard (count rate 
to flux) conversion factor (e.g., Kastner et al 2017); for {\it XMM} sources, the fluxes 
are those listed in the SSC. Bolometric fluxes used to estimate 
log($L_x$/$L_{bol}$) were obtained from a star's $G$ magnitude and 
spectral-type-dependent bolometric correction (Pecaut \& Mamajek 2013). The 
resulting NUV-G colors and log($L_x$/$L_{bol}$) values for the stars listed in 
Table 5 are within the ranges expected for stars of age $<$100 Myr (i.e., roughly 
-8 to -5 and -3.5 to -3.0, respectively; e.g., Kastner et al 2017). The lack of 
UV or X-ray counterparts to $\sim$2/3 of the stars in Tables 1 and 2 is readily 
explained as being due to the limited sensitivities of the {\it ROSAT} and {\it GALEX} all-sky 
surveys and the limited sky coverage of pointed {\it XMM-Newton} observations.

\section{Discussion} \label{sec:discussion}

At the outset, the primary goal of the present paper was to clarify the 
membership, structure, and age of the $\chi$$^1$ For cluster.  But, as is 
often the case in astronomy, arguably the most interesting finding of this 
research is a serendipitous discovery $-$ a remarkable concentration of 
M-type stars that possess excess mid-infrared emission (for details see later 
in this Section).  To the best of our knowledge, such a concentration is 
unprecedented in astronomy.

Following the procedure laid out in King (1962), we characterize the $\chi$$^1$ 
For cluster by a core radius and a limiting tidal radius even though neither of 
these concepts is quite appropriate for this cluster. King's core radius is 
calculated from a polynomial fit to stellar number counts vs. radial offset from 
the center of a richly populated (globular) cluster. Because the $\chi$$^1$ For 
cluster is far too sparse to apply this method, we simply denote the region in 
Figure \ref{fig:skyplot} with the largest stellar density, and a size of the order of a parsec, to be the core 
(Table 1). This radius (1 pc) encompasses more than half of the mass of the cluster
within the tidal radius.  This core region is far more densely packed, especially with stars of 
type A, F, and G, than  is the rest of the cluster (see discussion below). 

King argues that outer boundaries of old clusters are typically delineated by Galactic tides. But he 
recognized that the timescales for such tides to operate can be quite long, so 
King's analysis does not include young open clusters (such as $\chi$$^1$ 
For). Nonetheless, we calculate the tidal radius of the $\chi$$^1$ For cluster 
and consider all stars that have appropriate kinematic and photometric 
characteristics and that fall within this radius to be likely members of the 
cluster.  The tidal radius is equal to $R$($M_c$/3$M_g$)$^{1/3}$ (King 1962), where $R$, the distance between Earth and the Galactic Center, is equal to $\sim$8200 pc (Gravity Collaboration 2019), and $M_c$ and $M_g$ are, respectively, the total mass ($\sim$23 $M_\sun$) listed in Tables 1 and 2, and the mass of the Galaxy interior to the Sun.  For a solar circular rotation velocity of 240 km s$^{-1}$,  $M_g$ $\sim$10$^{11}$ $M_\sun$.  Then the tidal radius is $\sim$3.5 pc, the corresponding volume is $\sim$180 pc$^3$, and mass density is $\sim$0.1 $M_\sun$ pc$^{-3}$.  This is comparable to the local disk density (Mamajek 2016). 

Figure \ref{fig:cmd} is a {\it Gaia}-based CMD that includes likely $\chi$$^1$ 
For cluster members out to the tidal radius.  For an age of 40 Myr, the dividing line 
between stars and brown dwarfs lies near the absolute $G$ magnitude = 12.4 \citep{2015A&A...577A..42B}.
%(Baraffe et al 2015). 
Objects in Figure \ref{fig:cmd} and Tables 2 \& 3 near $M_G$ = 14 are brown dwarfs, while various other cluster members lie near the dividing line between stars and brown dwarfs.

\begin{figure}
\plotone{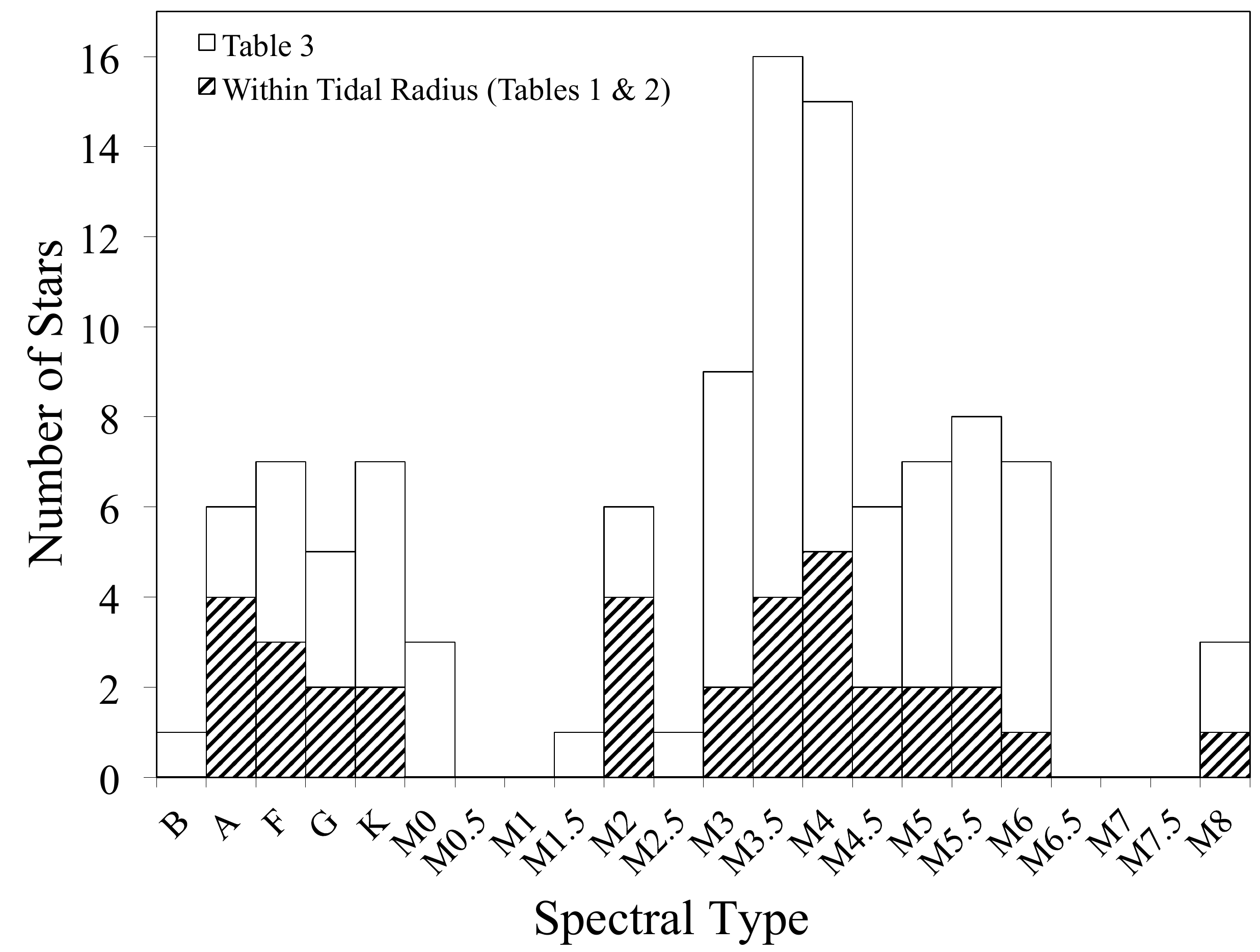}
\caption{Distribution of number of stars in $\chi$$^1$ For according to spectral type 
as counted from Tables 1$-$3.}
\label{fig:hist-sptype}
\end{figure}

Figure \ref{fig:hist-sptype} is a histogram of the number of cluster objects listed in Tables 1-3 
as a function of spectral 
type. It is apparent that the $\chi$$^1$ For cluster is dominated by mid-M type 
stars, as is typical of stars of various ages in the solar vicinity.  One may 
compare Figure \ref{fig:hist-sptype}  with the histograms for Tuc-Hor and Columba presented in Figure 9 
of Lee \& Song (2019). These histograms were constructed from studies before 
publication of the {\it Gaia} DR2 catalog, and thus should be relatively lacking in 
mid-to-late M-star members. We can anticipate, after the {\it Gaia} catalog has been 
exploited, that the Tuc-Hor and Columba histograms will look more like the one 
shown in Figure \ref{fig:hist-sptype}.

The mass segregation that is characteristic of clusters $-$ more massive stars preferentially 
near the cluster center $-$  is apparent in Tables 1 \& 2.  Of the 10 stars in the core of the $\chi$$^1$ For 
cluster (Table 1), seven are of spectral types AFG, while of the 24 stars between the core and tidal radii 
(Table 2), only two are of types AFG.

\begin{figure}
\plotone{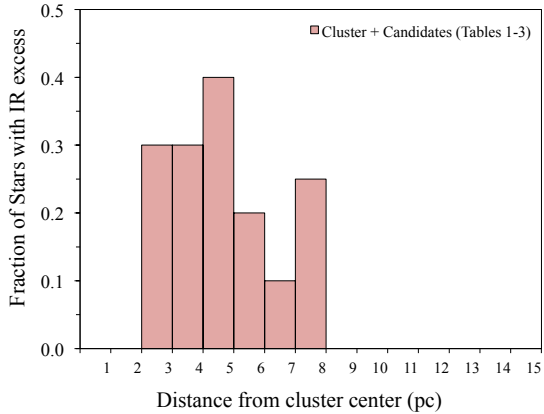}
\caption{Fraction of IR excess stars in Tables 1$-$3 as a function of distance from the center of the 
$\chi$$^1$ For cluster.}
\label{fig:hist-irxs}
\end{figure}

Dusty M-type stars are stunningly abundant in an 
annulus that surrounds the core of the $\chi$$^1$ For cluster (see Figure \ref{fig:hist-irxs}).  
Tables 1 \& 2 indicate 23 M-type stars within the tidal radius of the cluster.  Of these, 5 (or 22\%) display
infrared excess emission in the {\it WISE} catalog (Figure \ref{fig:sed}).  Alternatively, Tables 2 
\& 3 contain a total of 37 M-type stars in a three-dimensional annulus with inner 
and outer radii of 2 and 7 pc, respectively, centered on the mean position 
of the stars $\chi$$^1$ For and $\chi$$^3$ For.  Of these 37 stars, 10 (27\%) have 
excess IR emission.  And of the 10, most appear to include emission from 
dust with temperatures of at least a few 100 K.  These percentages 
are far higher than those obtained in all 
of the numerous searches that have been carried out for dust at M-type stars
aged 40 Myr or older  (see, e.g. Kraus et al.~2014; Binks \& Jeffries 2017; Flaherty et al.~2019, and references therein).

In contrast to the 27\% of dusty M-type stars between 2 and 7 pc from cluster center, between 7 and 15 pc there are 39 M-type stars of which none are dusty.  Put differently, between 0 and 7 pc 
there are 10 dusty M-type stars, while in a volume that is 9 times larger between 7 and 15 pc there are
none.  

M-type stars in the coeval Tuc-Hor Association lack warm dusty debris disks.  In a sample of $\sim$100 stars with spectral types between M0 and M6, Kraus et al (2014) found none that have excess emission in the {\it WISE}3 (11 $\mu$m) band $-$ this to be compared with the 8 in the annulus of dusty stars that surrounds $\chi$$^1$ For and $\chi$$^3$ For that possess excess W3 emission.

In Tables 2 \& 3, the occurrence of IR excesses at five stars that are members of wide binary systems (out of a total of 13 dusty stars) 
is consistent with the conclusion in Zuckerman (2015) and Silverberg et al (2018) 
that warm orbiting dust is more likely to be found at a young star that is a 
member of a wide binary system than at a single star. In addition to these wide 
binaries, one or both members of the relatively close (60 AU) 0.6" binary 
J0322-3547 is also dusty.  A plausible cause of the overabundance of dusty 
debris around members of binary stars is the eccentric Kozai-Lidov effect (Naoz 2016).  
Given an appropriate system age, the K-L effect can generate dynamical instabilities in planetary systems where the orbital plane of the system possesses a substantial inclination 
with respect to the orbital plane of a distant companion star.

The association of the $\chi$$^1$ For cluster with the Columba and Tuc-Hor 
Associations is based on similar UVW and similar locations on color-magnitude 
diagrams. Calculation of UVW requires knowledge of parallax, proper motion, and 
radial velocity. All stars in Tables 1-3 have parallax and proper motion 
consistent with cluster membership. Of the core stars in Table 1 only four 
($\chi$$^1$ For itself, $\chi$$^3$ For, CD-36 1289, and ChiFor 3) have 
measured radial velocities that we were able to find in the literature (however, see Note added in proof). 
We used 
these four stars to calculate the UVW given in the Notes for Table 1. These UVW 
(-12.3, -21.5, -4.6 km s$^{-1}$) may be compared with the UVW of
Columba (-12.2, -21.3, -5.6   km s$^{-1}$) and 
Tuc-Hor (-10.6, -21.0, -2.1  km s$^{-1}$) given in Mamajek (2016). The UVW of the massive B6 star 
i Eri (HD 24626) that is located outside of the tidal radius (see Figure \ref{fig:skyplot}
and below) is -13.3, -21.5, -5.1  km s$^{-1}$.

The stars in Tables 2 \& 3 with listed radial velocities can be divided into three groups.  The 
first is stars with UVW clearly consistent with the UVW of the $\chi$$^1$ For cluster given in
the previous paragraph.  A second group consists of stars for which the listed radial velocities, 
if correct, would likely mean that the star does not have the UVW of the cluster and thus is
not a member.  However, given the (large) size of the error bar of its radial velocity, the star could have the 
same UVW as the cluster.  The third group is composed of stars that are not cluster members
if the size of the radial velocity error bar is accurate.  Among the stars in Tables 2 \& 3, only HD 20854
and J0336-4231, and perhaps J0331-3358, are included in the third group.

Three well-studied clusters are located within $\sim$100 pc of Earth, two of which 
(the Hyades and Coma Berenices) are hundreds of millions of years old, while the 
age of the third ($\eta$ Cha) is $\sim$10 Myr (Mamajek 2016). Thus, at $\sim$40 
Myr, the $\chi$$^1$ For cluster is well positioned among the other nearby clusters 
for tracing the evolution of stars and substellar objects. The mass densities of 
the older clusters fall in the range 0.3-3.0 $M_\sun$ pc$^{-3}$, while the $\eta$ Cha 
core (containing 18 stars) is substantially denser at $\sim$30 $M_\sun$ pc$^{-3}$ 
(Mamajek 2016). As may be seen from Table 1, the core of the $\chi$$^1$ For 
cluster contains 10 stars (HD 21434 is an 0.2" binary) within a sphere of radius 1 pc 
at an average distance from 
Earth of 104 pc.  The mass density of the cluster core is about 3.0 $M_\sun$ pc$^{-3}$. 
This can be compared with the local disk density of $\sim$0.1 $M_\sun$ 
pc$^{-3}$ (Mamajek 2016). When, instead, the volume of the $\chi$$^1$ For
cluster out to its tidal radius of about 3.5 pc is considered, then the cluster mass 
density is only $\sim$0.1 $M_\sun$ pc$^{-3}$, i.e. comparable to the local disk density.

The two brightest and most massive core stars ($\chi$$^1$ For and $\chi$$^3$ 
For) would be just barely visible to a person 
with excellent eyesight in a dark sky site. A comoving, brighter, more massive, 
star i Eri is found outside of the cluster tidal radius (Figure \ref{fig:skyplot}). 
We have scrutinized the {\it Gaia} DR2 catalog for comoving stars located near i 
Eri.  Only two are present; these may be distant bound companions of
i Eri (see notes to Table 3). This relatively massive, obviously comoving, star 
has not been identified previously as a member of either the Columba or Tuc-Hor 
Associations; only one star ($\alpha$ Pav) more massive than i Eri has ever been 
suggested (Zuckerman \& Song 2004; Torres et al 2008) to be a member of these 
associations.

All young nearby stellar associations have been named after a constellation or an 
important stellar member of an association (Torres et al 2008; Mamajek 2016), or, 
in the case of naked eye star clusters, after Greek mythology. Given this precedent 
and the fact that there is only modest overlap between the few stars attributed to 
Alessi 13 (named after a person) in Dias et al (2002) and the stars listed in the 
tables in the present paper, we believe that $\chi$$^1$ For is a much more 
descriptive and appropriate name for this cluster. We hope the name $\chi$$^1$ For 
will be adopted in future work.

\section{Conclusions}

We have investigated the sparse, nearby, southern hemisphere open cluster 
$\chi$$^1$ For.  At a distance of $\sim$104 pc this cluster is sufficiently 
youthful and close to Earth that its entire main sequence and even the upper mass 
end of its brown dwarf sequence are accessible in the {\it Gaia} DR2 survey.  By 
comparison (in a {\it Gaia}-based CMD; Fig.~\ref{fig:cmd}) with the comoving and 
coeval Columba and Tucana-Horologium Associations, the age of the $\chi$$^1$ For 
cluster is seen to be $\sim$40 Myr.  The most massive stars within the core of the 
cluster are the A1-type stars $\chi$$^1$ For and $\chi$$^3$ For.  Perhaps 
surprisingly, a more massive star $-$ i Eri, spectral type B6 $-$ that may
be associated with this cluster, is located outside of the cluster 
tidal radius.

But perhaps the most notable finding of our study of the $\chi$$^1$
For cluster $-$ when compared to all previous studies of stars 
of age 40 Myr and older and located anywhere in the solar vicinity $-$ is an
unprecedented abundance of M-type stars 
that possess excess mid-infrared emission above that due to the
stellar photosphere. Based on their 
investigation of strong mid-IR emission, Melis et al (2010) deduced a 
characteristic age of 30-100 Myr for the epoch of final catastrophic mass 
accretion onto terrestrial planets around Sun-like stars.  The 40 Myr age of the 
$\chi$$^1$ For cluster and SEDs of many of its M-type 
members now suggest a similar rocky planet formation time scale around (the less 
massive) M-type stars. Consistent with past findings for dusty, field, M-type 
stars (Zuckerman 2015), wide binary systems are overrepresented in the entire $\chi$$^1$ For 
cluster infrared excess sample.  This involvement of a second massive body 
suggests a possible role to be played by the eccentric Kozai-Lidov effect (e.g., 
Naoz 2016), in dynamically stirring up planetesimal belts. The
location of IR excess stars within a well-defined annular region around the cluster
core also hints at the influence of stellar neighbors on the stability of planetary
systems.

Based on its location in the plane of the sky, its age, and its Galactic space 
motions UVW, the $\chi$$^1$ For cluster is plausibly intimately associated with 
one or both of the Columba and Tucana-Horologium Associations.  But all existing 
studies of these two spatially dispersed moving groups are limited to stars with 
typical distances from Earth of well less than 100 pc. Future {\it Gaia}-based studies 
should reveal whether or not these two associations physically surround the 
$\chi$$^1$ For cluster.

\subsection*{Note added in proof}  As this paper was going to press, A. Moor (private communication) provided RV and lithium (Li)
equivalent width (EW) measurements for a core $\chi$$^1$ For member (CD-36 1309) and a candidate member 
(CD-37 1224) that confirm their status as members of the cluster (see Table 6).

\begin{table}[h]
\caption{$\chi$$^1$ For Members: New Data }
\begin{center}
\begin{tabular}{lccc}
\hline 
\hline
Star & RV & U,V,W & EW (Li)   	\\
          & (km/s)  & (km/s) &  (m\AA)  \\
\hline
CD-36 1309 & 20.2$\pm$2.0  & -13.1,-23.2,-6.3 & 170$\pm$15 \\
CD-37 1224 & 17.4$\pm$2.2  & -12.2,-21.7,-5.0 &  225$\pm$10 \\
\hline
\end{tabular}
\end{center}
\end{table}

\subsection*{Acknowledgements}
We thank the referee for a timely report and suggestions that helped to improve the paper.  This research was supported in part by grants to UCLA from NASA. J.H.K.'s research on young stars near Earth is supported by NASA Astrophysics Data Analysis Program grant NNX12AH37G and NASA Exoplanets Research Program grants NNX16AB43G and 80NSSC19K0292 to RIT.

\clearpage


\begin{thebibliography}{}

\bibitem[Baraffe et al.(2015)]{2015A&A...577A..42B} Baraffe, I., Homeier, D., Allard, F., et al.\ 2015, \aap, 577, A42
\bibitem[Bianchi et al.(2017)]{2017ApJS..230...24B} Bianchi, L., Shiao, B., \& Thilker, D.\ 2017, \apjs, 230, 24B 
\bibitem[Binks, \& Jeffries(2017)]{2017MNRAS.469..579B} Binks, A.~S., \& Jeffries, R.~D.\ 2017, \mnras, 469, 579
\bibitem[Cantat-Gaudin et al.(2018)]{2018A&A...618A..93C} Cantat-Gaudin, T., Jordi, C., Vallenari, A., et al.\ 2018, \aap, 618, A93
\bibitem[de la Reza et al.(1989)]{1989ApJ...343..L61D} de la Reza, R., Torres, C. A. O., Quast, G., et al.\ 1989,\apj, 343, L61
\bibitem[Dias et al.(2002)]{2002A&A...389..871D} Dias, W.~S., Alessi, B.~S., Moitinho, A., et al.\ 2002, \aap, 389, 871
\bibitem[Dufour et al.(2017)]{2017ASPC..509....3D} Dufour, P., Blouin, S., Coutu, S., et al.\ 2017, The Montreal White Dwarf Database, 20th European White Dwarf Workshop, (San Francisco, CA, Astr. Soc. of the Pacific), p. 3
\bibitem[Flaherty et al.(2019)]{2019ApJ...872...92F} Flaherty, K., Hughes, A.~M., Mamajek, E.~E., et al.\ 2019, \apj, 872, 92
\bibitem[Gagn{\'e} et al.(2018)]{2018ApJ...865..136G} Gagn{\'e}, J., Faherty, J.~K., \& Mamajek, E.~E.\ 2018, \apj, 865, 136
\bibitem[{\it Gaia} Collaboration et al.(2018)]{2018A&A...616A...1G} {\it Gaia} Collaboration, Brown, A.~G.~A., Vallenari, A., et al.\ 2018, \aap, 616, A1
\bibitem[Gentile Fusillo et al.(2019)]{2019yCat..74824570G} Gentile Fusillo, N.~P., Tremblay, P.-E., Gaensicke, B.~T., et al.\ 2019, \mnras, 482, 4570
\bibitem[Gravity Collaboration et al.(2019)]{2019A&A...625L..10G} Gravity Collaboration, Abuter, R., Amorim, A., et al.\ 2019, \aap, 625, L10
\bibitem[Gregorio-Hetem et al.(1992)]{1992AJ....103..549G} Gregorio-Hetem, J., Lepine, J. R. D., Quast, G. et al. \ 1992 \aj, 103, 549
\bibitem[Kastner et al.(2017)]{2017ApJ...841...73K} Kastner, J.~H., Sacco, G., Rodriguez, D., et al.\ 2017, \apj, 841, 73
\bibitem[Kastner et al.(1997)]{1997Sci...277...67K} Kastner, J.~H., Zuckerman, B., Weintraub, D.~A., et al.\ 1997, Science, 277, 67
\bibitem[Kharchenko et al.(2016)]{2016A&A...585A.101K} Kharchenko, N.~V., Piskunov, A.~E., Schilbach, E., et al.\ 2016, \aap, 585, A101
\bibitem[King(1962)]{1962AJ.....67..471K} King, I.\ 1962, \aj, 67, 471
\bibitem[Kraus et al.(2014)]{2014AJ....147..146K} Kraus, A.~L., Shkolnik, E.~L., Allers, K.~N., et al.\ 2014, \aj, 147, 146
\bibitem[Lee, \& Song(2019)]{2019MNRAS.486.3434L} Lee, J., \& Song, I.\ 2019, \mnras, 486, 3434
\bibitem[Lindegren et al.(2018)]{2018A&A...616A...2L} Lindegren, L., Hern{\'a}ndez, J., Bombrun, A., et al.\ 2018, \aap, 616, A2
\bibitem[Lodieu et al.(2019)]{2019arXiv190603924L} Lodieu, N., Perez-Garrido, A., Smart, R.~L., et al.\ 2019, A\&A, 628, A66
\bibitem[Mamajek(2016)]{2016IAUS..314...21M} Mamajek, E.~E.\ 2016,  in
  Proceedings of  IAU Symp. 314, {\it Young Stars \& Planets Near the Sun}, eds. J. H. Kastner et al. (Cambridge: Cambridge Univ. Press) 21
\bibitem[Mason et al.(2015)]{2015yCat....102026M} Mason, B.~D., Wycoff, G.~L., Hartkopf, W.~I., et al.\ 2015, VizieR Online Data Catalog, B/wds
\bibitem[Melis et al.(2010)]{2010ApJ...717L..57M} Melis, C., Zuckerman, B., Rhee, J.~H., et al.\ 2010, \apjl, 717, L57
\bibitem[Naoz(2016)]{2016ARA&A..54..441N} Naoz, S.\ 2016, \araa, 54, 441
\bibitem[Pecaut, \& Mamajek(2013)]{2013ApJS..208....9P} Pecaut, M.~J., \& Mamajek, E.~E.\ 2013, \apjs, 208, 9
\bibitem[P\"ohnl \& Paunzen (2010)]{2010A&A...514A..81P} P\"ohnl, H. \& Paunzen, E.\ 2010, A\&A, 514A, 81P 
\bibitem[Silverberg et al.(2018)]{2018ApJ...868...43S} Silverberg, S.~M., Kuchner, M.~J., Wisniewski, J.~P., et al.\ 2018, \apj, 868, 43
\bibitem[Torres et al.(2008)]{2008hsf2.book..757T} Torres, C.~A.~O., Quast, G.~R., Melo, C.~H.~F., et al.\ 2008, Handbook of Star Forming Regions, Volume 2, (San Francisco, CA, Astr. Soc. of the Pacific), p. 757
\bibitem[Wright et al.(2010)]{2010AJ....140.1868W} Wright, E.~L., Eisenhardt, P.~R.~M., Mainzer, A.~K., et al.\ 2010, \aj, 140, 1868
\bibitem[Yen et al.(2018)]{2018A&A...615A..12Y} Yen, S.~X., Reffert, S., Schilbach, E., et al.\ 2018, \aap, 615, A12
\bibitem[Zuckerman(2015)]{2015ApJ...798...86Z} Zuckerman, B.\ 2015, \apj, 798, 86
%\bibitem[Zuckerman(2019)]{2019ApJ...870...27Z} Zuckerman, B.\ 2019, \apj, 870, 27
\bibitem[Zuckerman, \& Song(2004)]{2004ARA&A..42..685Z} Zuckerman, B., \& Song, I.\ 2004, \araa, 42, 685
\bibitem[Zuckerman et al.(2001)]{2001ApJ...559..388Z} Zuckerman, B., Song, I., \& Webb, R.~A.\ 2001, \apj, 559, 388


\end{thebibliography}
\end{document}